\documentclass[prl,twocolumn,superscriptaddress]{revtex4}

\usepackage{graphicx}
\usepackage{dcolumn}
\usepackage{bm}
\usepackage{natbib}
\usepackage{amsmath}
\usepackage{here}

\begin{document}
\preprint{preprint - not for distribution}

\title{Controlled growth of InAs nanowires on engineered substrates}

\author{Stefano Roddaro}
\affiliation{NEST CNR-INFM and Scuola Normale Superiore, P.za S.Silvestro 12, 56127 Pisa, Italy}

\author{Philippe Caroff}
\affiliation{Solid State Physics \& the Nanometer Structure 
 Consortium, Lund University, P.O. Box 118 Lund, Sweden}

\author{Giorgio Biasiol}
\affiliation{TASC CNR-INFM Laboratory, Area Science Park, 34012 Trieste, Italy}

\author{Francesca Rossi}
\affiliation{Istituto CNR-IMEM, Parco Area delle Scienze 37/A, 43100 Parma, Italy}

\author{Claudio Bocchi}
\affiliation{Istituto CNR-IMEM, Parco Area delle Scienze 37/A, 43100 Parma, Italy}

\author{Kristian Nilsson}
\affiliation{Solid State Physics \& the Nanometer Structure 
 Consortium, Lund University, P.O. Box 118 Lund, Sweden}
 
\author{Linus Fr\"oberg}
\affiliation{Solid State Physics \& the Nanometer Structure 
 Consortium, Lund University, P.O. Box 118 Lund, Sweden}
 
\author{Jakob B. Wagner}
\affiliation{Solid State Physics \& the Nanometer Structure 
 Consortium, Lund University, P.O. Box 118 Lund, Sweden}
 
\author{Lars Samuelson}
\affiliation{Solid State Physics \& the Nanometer Structure 
 Consortium, Lund University, P.O. Box 118 Lund, Sweden}

\author{Lars-Erik Wernersson}
\affiliation{Solid State Physics \& the Nanometer Structure 
 Consortium, Lund University, P.O. Box 118 Lund, Sweden}

\author{Lucia Sorba}
\affiliation{NEST CNR-INFM and Scuola Normale Superiore, P.za S.Silvestro 12, 56127 Pisa, Italy}
\affiliation{TASC CNR-INFM Laboratory, Area Science Park, 34012 Trieste, Italy}
 
\date{\today}

\begin{abstract}
We demonstrate the Au-assisted growth of semiconductor nanowires on different engineered substrates. Two relevant cases are investigated: GaAs/AlGaAs heterostructures capped by a $50\,{\rm nm}$-thick InAs layer grown by molecular beam epitaxy and a $2\,{\rm \mu m}$-thick InAs buffer layer on Si(111) obtained by vapor phase epitaxy. Morphological and structural properties of substrates and nanowires are analyzed by atomic force and transmission electron microscopy. Our results indicate a promising direction for the integration of III-V nanostructures on Si-based electronics as well as for the development of novel micromechanical structures.
\end{abstract}

\maketitle

\section{I. INTRODUCTION}

Recent years have witnessed the emergence of semiconductor nanowires (NWs) as a new promising platform for nano- and opto-electronics~\cite{NanoE1,NanoE2}. The small radial extension of the NW structure reduces the strain constraint and allows for novel heterostructure combinations~\cite{InAsInP,InAsInSb,InAsNWonGaAs,InAsInPRadial}. The strong surface effects during NW growth can lead to high-quality nanostructures from both the crystalline and geometrical point of view~\cite{Quality1}. While the vast majority of the present fabrication strategies relies on random deposition processes and marker-aligned metallization of contact electrodes~\cite{Planar}, the achievement of a production-oriented NW technology will most likely require a different approach. An interesting possibility consists in the fabrication of vertical devices starting from NWs grown at controlled positions on the surface~\cite{WigFET,CV}. The successful development of complex circuits based on vertical NWs, however, will require the development of suitable substrates in order to achieve -- for instance -- a simple independent and flexible addressing of different NW elements grown on the same chip. More generally, the growth of NWs starting from complex buffer layers or even patterned structures is still in need of a more exhaustive evaluation.

In this paper, we demonstrate the Au-seeded growth of InAs NWs starting from specifically engineered substrates. Two cases are analyzed in Section II and III, respectively: (i) NW growth on a thin InAs layer deposited on top of a GaAs/AlGaAs engineered substrate realized by molecular beam epitaxy (MBE); (ii) NW growth on a $2\,{\rm \mu m}$-thick InAs buffer layer deposited a on Si(111) substrate by metal-organic vapor phase epitaxy (MOVPE). While GaAs/AlGaAs substrates are interesting in view of the integration of NWs in high-mobility systems, in the second case we aim at evaluating a possible integration path for III-V nanostructures in Si-based platforms.  Both approaches address complex challenges for the substrate engineering such as the growth of buffer layers on (111) III/V substrates and the formation of antiphase domains on Si.

\section{II. RESULTS AND DISCUSSION: InAs NWs on InAs/AlGaAs (111)}

GaAs/AlGaAs epitaxy is a well-developed technology that can be used to fabricate complex lattice-matched layered structures and buffer layers are widely used for the growth of high-In containing InGaAs alloys on GaAs substrates. As InAs NWs preferentially grow on the (111)B substrates, the established technology must be adopted to the different substrate orientation. The buffer layer technology is attractive for NWs as heterostructured substrates can also be useful to define thin conducting layers on top of insulating substrates. This is crucial in order to address individual devices in a complicated multi-NW circuit and to reduce stray capacitive couplings. In addition, ${\rm Al_xGa_{1-x}As}$ alloys can have a strong etching selectivity as a function of the composition parameter (x) and they can be used to fabricated free-standing micromechanical structures such as membranes and cantilevers~\cite{FreeStandingAlGaAs, NWCantilevers}. 

The direct nucleation of InAs NWs on GaAs is challenging~\cite{InAsNWonGaAs} and the properties of (111) InAs/GaAs heterojunctions are still largely unexplored. For these reasons we developed a growth procedure to obtain a thin, doped and high-quality InAs layer on top of semi-insulating GaAs/AlGaAs heterostructures. Given the high lattice mismatch ($\approx 7\%$) between (Al,Ga)As alloys and InAs, we opted for the insertion of an ${\rm In_xGa_{1-x}As}$ buffer layer (BL) in the deposition sequence in order to obtain a high-quality InAs top layer~\cite{Capotondi}. 

\begin{figure}[ht!]
\begin{center}
\includegraphics[width=0.48\textwidth]{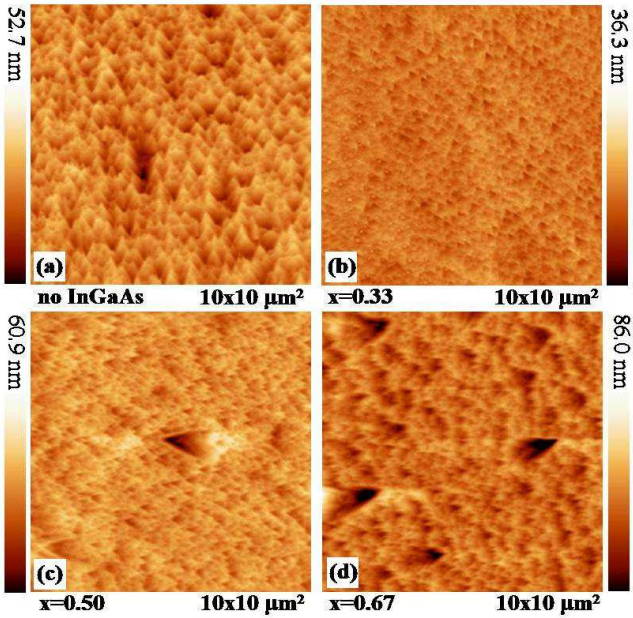}
\caption{$10\times 10\,{\rm \mu m^2}$ AFM scans of $50\,{\rm nm}$-thick InAs layers grown directly on GaAs(111)B (sample $a$), or after the insertion of a $25\,{\rm nm}$ ${\rm In_xGa_{1-x}As}$ BL with ${\rm x}=0.33$ (sample $b$, see also Fig.~2a), ${\rm x}=0.5$ (sample $c$) and ${\rm x}=0.67$ (sample $d$).}
\end{center}
\end{figure}

InAs/AlGaAs structures were grown by solid source MBE on GaAs(111)B substrates misoriented by $3^\circ$ towards the $\left[ 2\,\overline{1}\,\overline{1}\right]$ direction. The layer sequence was chosen in order to be suitable for the fabrication of free-standing micromechanical structures by selective etching (see supplementary material). Starting from a GaAs(111)B substrate, a $250\,{\rm nm}$-thick ${\rm Al_{0.75}Ga_{0.25}As}$ stop-etch layer was grown at $630\,{\rm^\circ\!C}$ followed by $1\,{\rm\mu m}$ of GaAs grown at $610\,{\rm^\circ\!C}$. The resulting top surface had a typical RMS roughness of $\approx 0.6\,{\rm nm}$. The InAs cap layer was optimized in terms of strain relaxation, dislocation density and surface roughness. Deposition temperature was selected to be $300\,{\rm^\circ\!C}$, as a trade-off between crystal quality and surface flatness~\cite{Hooper}; growth rate and V/III ratio did not have significant effects on the roughness and were selected to be $0.1\,{\rm nm/sec}$ and $30$ (beam equivalent pressure ratio), respectively. The final InAs thickness was $50\,{\rm nm}$, since a residual presence of Ga atoms migrating from the substrate was detected at the surface in thinner layers~\cite{Wen}, while in thicker layers we observed an increase of the surface roughness. Figure~1 shows a series of atomic force microscopy (AFM) topographic pictures of the top InAs surface for samples with different ${\rm In_xGa_{1-x}As}$ BLs. InAs in sample $a$ was grown directly on GaAs and exhibits a surface roughness of $\approx6.1\,{\rm nm}$ RMS. Such a roughness can be significantly reduced (RMS $3.7\,{\rm nm}$) by the insertion of a $25\,{\rm nm}$ BL with ${\rm x}=0.33$ (sample $b$). We observed that increasing the In content of the BL causes the formation of $\approx50\,{\rm nm}$-deep arrowhead-shaped holes, as it can be seen in samples $c$ (${\rm x} = 0.5$) and $d$ (${\rm x}=0.67$), respectively. 

\begin{figure}[h!]
\begin{center}
\includegraphics[width=0.48\textwidth]{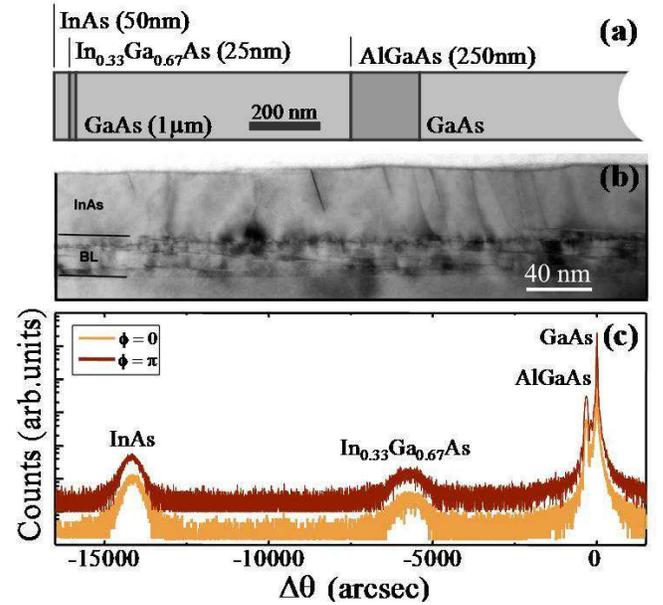}
\caption{(a) The layer sequence in the InAs/(Al,Ga)As substrate (sample $b$). (b) Bright field [110] cross-sectional TEM image of sample B: defects are located mainly in the BL. In the InAs layer, some threading dislocations are evident as dark segments preferentially along the (211) direction which results from (111) planes in the (110) projection. (c) XRD Bragg scan of sample $b$ along two azimuths rotated by $180^\circ$.}
\end{center}
\end{figure}

Given the lower roughness, sample $b$ (see layer sequence in Fig.~2a) was choosen for the investigation of NW growth and studied by bright field transmission electron microscopy (TEM). Figure~2b reports an image taken near [110] zone axis in (111) diffraction conditions and demonstrates that defects due to strain are mainly located at the BL. Strain relaxation was assessed by high-resolution X-ray diffraction (XRD) on symmetric (333) reflection at two different azimuths rotated by $180^\circ$. Despite the $3^\circ$ miscut of the GaAs(111)B substrate, Bragg peak positions are virtually identical at the two azimuthal angles, indicating a complete strain relaxation both in the ${\rm In_{0.33}Ga_{0.67}As}$ buffer layer and in the InAs top layer. Following van-der-Pauw measurements, the intentionally-doped InAs cap layer was found to have a carrier density and mobility of $n=1.7\times 10^{18}\,{\rm cm^{-3}}$ and $\mu=4.5\times10^3\,{\rm cm^2/Vs}$, respectively.

\begin{figure}[ht!]
\begin{center}
\includegraphics[width=0.48\textwidth]{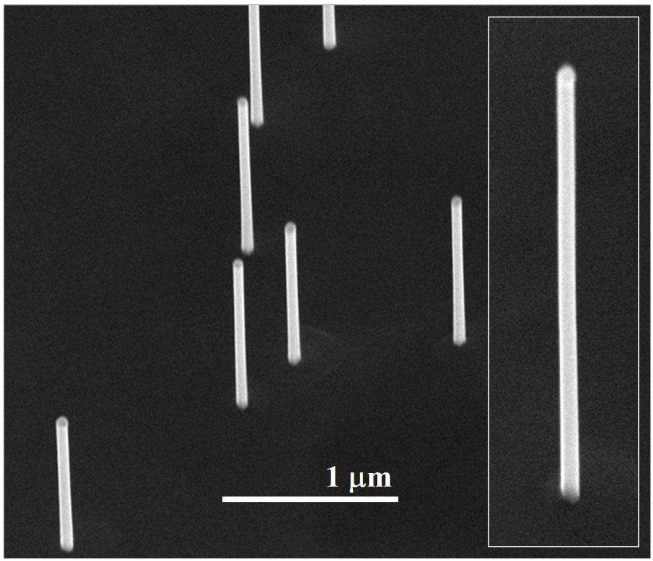}
\caption{Main panel: SEM image (tilt $15^\circ$) of InAs NWs grown on an InAs/(Al,Ga)As engineered substrates starting from Au nanoparticles deposited by aerosol technology. Inset: magnification of of a single NW.}
\end{center}
\end{figure}

InAs NWs were fabricated by means of chemical beam epitaxy (CBE) seeded by Au aerosol nanoparticles~\cite{NWGrowthRef}. Substrates were first annealed and deoxidized at $520\,{\rm ^\circ\!C}$ for $17$ minutes under As pressure; NW were then grown at $425^\circ\,{\rm C}$ for $60$ minutes using trimethylindium (TMIn) and pre-cracked tertiarybutylarsine (TBAs) with respective partial pressures of $0.15$ and $1.5\,{\rm mbar}$ (supply line) as precursors~\cite{NWGrowthRef}. Figure~3 demonstrates the successful nanoparticle-seeded deposition of NWs on sample $b$ (SEM image at $15^\circ$ tilt). NWs grow normal to the sample surface similarly to what observed on bulk InAs(111)B substrates. NWs had a diameter of $\approx60\,{\rm nm}$ and the right inset reports a higher magnification view of a single wire. A statistical analysis of the NW length gives an average value of about $2.8\pm 0.4\,{\rm \mu m}$, which is $\approx 25\%$ shorter compared to the one of NWs grown in parallel on a standard InAs substrate ($3.7\pm0.2\,{\rm \mu m}$). Such a difference might be linked either to a different surface mobility for In on the InAs(111)B cap layer of sample B in comparison with a bulk InAs(111)B substrate or to difficulties in the nucleation process. Despite this, our study shows these substrates are well-suited for NW growth.

\section{III. RESULTS AND DISCUSSION:  InAs NWs on InAs/Si(111)}

\begin{figure}[h!]
\begin{center}
\includegraphics[width=0.48\textwidth]{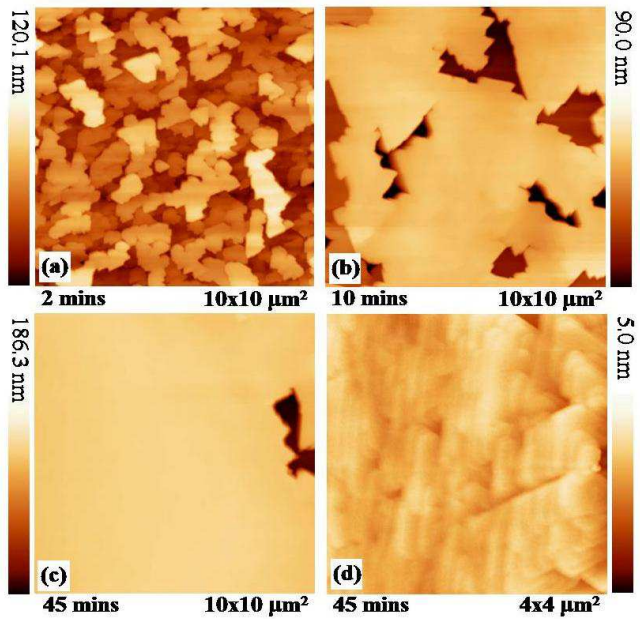}
\caption{AFM images of the surface after the second growth step as a function of growth time: (a) 1 minute, (b) 10 minutes, (c) 45 minutes and (d) zoom-in of panel (c). RMS in the last scan is $2.2\,{\rm nm}$.}
\end{center}
\end{figure}

The growth of III-V semiconductors on Si has attracted significant research efforts in the past decades as it holds the promise of a straight-forward integration of high-mobility and optically-active materials on top of the mainstream low-cost Si-based electronics. Due to the extreme difficulties linked to the high lattice mismatch between these materials however very few studies have been reported on the deposition of InAs on top of a Si substrate because of severe issues such as the formation of dense dislocations networks, thermal cracks and anti-phase domains (APD)~\cite{APB}. While cracks can be easily identified by optical microscopy, APDs are not so easy to detect and can form as a consequence of growth of polar III-V material on top of the non-polar Si surface. In the case of a Si(111) substrate, resulting III-V layers can contain at the same time some portions of surface oriented in the (111)A and others in the (111)B direction. As demonstrated in the following paragraphs, our epitaxial procedure preferentially develops (111)B-oriented domains and InAs NWs grow in the vertical direction.

InAs buffer layers were deposited directly on Si(111) substrates by MOVPE using a two-step procedure described in detail in Ref.~\cite{PhilippeInAsSi}. After the growth of a first nucleation layer of Stranski-Krastanow islands, a second growth step was used to reconstruct a flat layer of good morphological and conducting properties. Figures~4a-d display the surface structure as a function of the growth time during the second step. AFM images indicate that the InAs layer grows via a mechanism of triangular nuclei extension, probably limited by the diffusion length of the group III ad-atoms on the surface~\cite{PhilippeInAsSi}. For InAs layers thicker than $\approx1\,{\rm \mu m}$, the surface is characterized by large regions of small roughness ($2.2\,{\rm nm}$), even if portions with larger roughness could not be completely eliminated (see right side of Fig.~4c). The InAs layer on which NWs were grown was obtained by a $180$ minutes deposition and had a total thickness of approximately $2\,{\rm \mu m}$. This unintentionally doped substrate was also characterized by van-der-Pauw measurements. Carrier density and mobility at room temperature was measured to be $n=9\times10^{16}\,{\rm cm^{-3}}$ and $\mu\approx 5\times10^3\,{\rm cm^2/Vs}$, respectively.
 
\begin{figure}[t!]
\begin{center}
\includegraphics[width=0.48\textwidth]{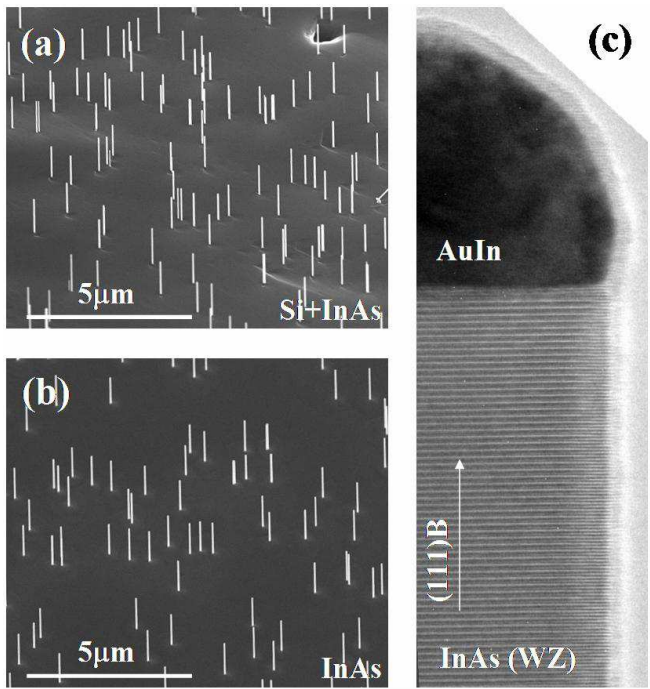}
\caption{(a) SEM image of InAs NWs grown on a thick InAs film deposited on Si(111). (b) SEM image of InAs NWs grown on a standard InAs(111)B substrate. Both images were taken at a tilt angle of $30^\circ$. (c) High-resolution TEM image of an InAs NW grown on a InAs/Si substrate.}
\end{center}
\end{figure}

Figure~5a shows a $30^\circ$-tilted SEM image of InAs NWs grown on the $2\,{\rm \mu m}$-thick InAs/Si substrate following the same procedure indicated in the previous section, but with a shorter deposition time ($30$ minutes). The growth process is normal and the NWs had a length of $1.7\pm 0.15\,{\rm \mu m}$ and a diameter of $\approx65\,{\rm nm}$. This can be compared with the length of $1.7\pm 0.1\,{\rm \mu m}$ obtained on a control InAs(111)B substrate (see Fig.~5b). The negligible statistical difference between the two cases indicates that these engineered substrates can be successfully employed for NW growth. We note that growth of InAs NWs on these layers provides an original tool to detect APDs. Indeed, NWs can only grow vertically with such a high yield on the (111)B surface while one would expect non-vertical growth directions and different growth rates on a (111)A surface~\cite{APDDetection}. The comparable growth rate and the perfect vertical orientation for InAs/Si(111) and InAs/InAs(111)B samples on large areas demonstrate the almost perfect (111)B orientation of the substrate (Fig.~5a,b). Growth starting from lithographically-defined Au nanoparticles was also investigated and gave good yield of NW arrays (see supplementary material and reference~\cite{WigFET} for applications to wrap-gate transistors). In order to gain insight about the quality of the crystal structure of NWs grown on the InAs/Si substrates, high resolution TEM evaluation was performed and compared to the state-of-the-art homoepitaxial InAs NWs grown by CBE. Figure~5c demonstrates that NWs crystallize in the usual hexagonal wurtzite phase, with nearly no stacking faults (density $\ll1\,{\rm \mu m^{-2}}$). It can be concluded that the engineered substrate does not affect in any significant way the nucleation nor the crystal quality of the NWs. 

It is worth noting that different approaches for the integration of III-V NWs on Si have been developed in recent times: one example is the direct growth of InAs NWs on Si, a nice demonstration of the flexibility offered by the NW growth technique~\cite{NWonSi,FETonSi}. The alternative strategy analyzed in this work presents more lattice matching issues but also significant advantages: it eliminates problems related to undesired InAs/Si heterojunctions at the base of the NW, i.e. directly in the conduction path; it provides a natural connection layer that can be patterned to address single devices in a multi-NW chip. In addition to this, other groups demonstrated the growth from {\em peeled} thin InAs(111)B layers~\cite{Peel}: while such a strategy clearly yields high-quality InAs cap layers, it relies critically on the perfect flatness of the substrate and cannot be used with patterned substrates.

\section{IV. Conclusions}

We have demonstrated the controlled growth of InAs NWs starting from two different engineered substrates: a $50\,{\rm nm}$-thick InAs cap layer grown on top of a GaAs/AlGaAs heterostructure and a $2\,{\rm \mu m}$-thick InAs buffer layer deposited on a Si(111) wafer. Both substrates offer specific advantages with respect to bulk InAs(111)B. In both cases the top layers can be patterned and allow the selective addressing of single NWs as well as the reduction of stray capacitances to the electrodes in wrap-gate transistor design. Finally, we demonstrated the growth of NWs starting from a GaAs/AlGaAs high-mobility structure designed for the realization of micromechanical structures by selective etching.

This work was supported in part by the Italian Ministery of University and Research through the FIRB project RBIN067A39. Also, this work was supported in part by the Swedish Foundation for Strategic Research (SSF), by the Knut and Alice Wallenberg Foundation and by the Swedish Research Council (VR).

\end{document}